\documentclass[amsmath,amssymb,showkeys,superscriptaddress,floatfix,aps,prd,10pt]{revtex4}
\usepackage{graphicx}
\usepackage{pdfpages}
\usepackage[caption=false]{subfig}
\usepackage{multirow}
\usepackage{appendix}
\usepackage{xcolor}
\usepackage[colorlinks=true, urlcolor=blue, linkcolor=blue, citecolor=green]{hyperref}
\def\pslash{p\!\!\!\slash }
\def\qslash{q\!\!\!\slash }
\def\xslash{x\!\!\!\slash }

\def\pslash{p\!\!\!\slash }
\def\qslash{q\!\!\!\slash }
\def\xslash{x\!\!\!\slash }

\def\beq{\begin{equation}}
\def\eeq{\end{equation}}
\def\bea{\begin{eqnarray}}
\def\eea{\end{eqnarray}}
\def\beeq{\begin{eqnarray}}
\def\eeeq{\end{eqnarray}}

\def\vel{\left|}
\def\ver{\right|}
\def\nnb{\nonumber}

\def\nnb{\nonumber}

\def\ba{\begin{array}}
\def\ea{\end{array}}

\def\xis0{{\Xi^{*0}}}

\def\g5{\gamma_5}

\def\es{\!\!\! &=& \!\!\!}

\def\ar{&+& \!\!\!}
\def\ek{&-& \!\!\!}

\begin{document}

\title{Gravitational form factors of $N(1535)$ in QCD }

\author{K. Azizi}
\affiliation{Department of Physics, University of Tehran, North Karegar Avenue, Tehran
14395-547, Iran}
\affiliation{Department of Physics, Dogus University, Acibadem-Kadikoy, 34722 
Istanbul, Turkey}
\author{U. \"{O}zdem}
\affiliation{Health Services Vocational School of Higher Education, Istanbul Aydin University, Sefakoy-Kucukcekmece, 34295 Istanbul, Turkey}

\begin{abstract}
We calculate the gravitational form factors of the excited  $N(1535)$ state with the quantum numbers $I(J^P)=\frac{1}{2}(\frac{1}{2}^-)$ via light cone QCD sum rules (LCSR).  To this end, we  consider the quark part of the  energy-momentum tensor (EMT) current   and use the general form of the nucleon's interpolating field as well as the distribution amplitudes (DAs) of $N(1535)$.  As both the nucleon and $N(1535)$ couple to the same current, the $N(1535) \rightarrow N$ gravitational transition form factors are entered to the calculations as the main input parameters.  First we revisit the transitional gravitational form factors of  $N(1535) \rightarrow N$, then extract the values of the form factors of the  $N(1535)$ excited state. We observe that the gravitational  form factors of $N(1535)$ in terms of $Q^2$ are well described by the  multipole fit function. As a byproduct, we also calculate the  pressure and  energy density at the center of $N(1535)$ and estimate its mechanical radius.
\end{abstract}
 \date{\today}
\maketitle

\section{Introduction} 
The form factors  (FFs) are fundamental parameters that contain useful information on the nature, internal structure and geometric shapes of the hadrons. By calculation of electromagnetic, axial, tensor and gravitational or energy momentum form factors, one can get valuable information on the distribution of charge,  magnetism, pressure, energy and other mechanical observables  inside the hadrons. Among these form factors, the energy-momentum tensor form factors (EMTFFs) or gravitational form factors (GFFs) play essential roles as they are also sources of knowledge on the fractions of the momenta carried by the quarks and gluons as ingredients of the composite objects.  They carry information on the  total angular momenta of quarks and gluons as well as  distribution and stabilization of the strong force inside the hadrons, as well.  The GFFs of hadrons are in the focus of much attention. By recent experimental progresses on the gravitational form factors of nucleons \cite{Burkert:2018bqq,Burkert:2019kxy}, we hope it will be possible to experimentally investigate the excited nucleons and other light and heavy baryons to measure their GFFs and those defining the transitions among the hadrons. Theoretical and phenomenological investigations on various  GFFs of different hadrons can play essential role in the present time as they can help experimental groups in measuring GFFs. 
For the first time, the GFFs were introduced  by Kobzarev and Okun
 in 1962 \cite{Kobzarev:1962wt}.  Then the GFFs of nucleons were calculated  using different  methods and  approaches: chiral perturbation theory ($\chi$PT) \cite{chen:2001pva, Ando:2006sk, Diehl:2006ya,Belitsky:2002jp,Diehl:2006js, Dorati:2007bk},  Bag model~\cite{Neubelt:2019sou}, instanton picture (IP)~\cite{Polyakov:2018exb},  chiral quark soliton model ($\chi$QSM) \cite{Wakamatsu:2007uc,Petrov:1998kf, Schweitzer:2002nm, Ossmann:2004bp, Wakamatsu:2005vk,  Goeke:2007fq,Goeke:2007fp, Jung:2013bya,Jung:2014jja, Jung:2015piw,Wakamatsu:2006dy}, dispersion relation (DR)~\cite{Pasquini:2014vua}, 
Skyrme model \cite{Cebulla:2007ei,Kim:2012ts}, lattice QCD \cite{Gockeler:2003jfa, Bratt:2010jn, Hagler:2007xi,Brommel:2007sb,Hagler:2003jd,mathur:1999uf,Negele:2004iu,Deka:2013zha},  LCSR~\cite{Anikin:2019kwi,Azizi:2019ytx}  and   instant-front form (IFF) \cite{Lorce:2018egm}. The GFFs of hyperons, for the first time, were calculated in \cite{Ozdem:2020ieh}.

In the present study, we calculate the gravitational form factors of the excited  $N(1535)$ (hereafter  $N^*$) state with the quantum numbers $I(J^P)=\frac{1}{2}(\frac{1}{2}^-)$ via LCSR by  considering the quark part of the  EMT current   and using the general form of the nucleon's interpolating field as well as the DAs of $N^*$  \cite{maxiphd} .  As both the nucleon and $N^*$ couple to the same current, the $N^* \rightarrow N$ gravitational transition form factors enter to the calculations as the main input parameters.  First, we revisit  the  transition GFFs of $N^*\rightarrow N $  studied  in Ref. \cite{Ozdem:2019pkg}. In this work,  the transition matrix element of the energy momentum tensor (EMT) current is parameterized in terms of four independent FFs. Considering the conservation of  EMT current  and the Lorentz invariance   as well as  avoiding from the redundancy in Lorentz structures at the limit of equal masses  of the initial and final baryonic states, one can parameterize the transition matrix element  of the total EMT current, sandwiched between the N and   $N^*$ states  in terms of three  different form factors (for details see \cite{Polyakov:2020rzq}). If one applies the partial EMT current (say the quark part), which is not conserved, one has to add three more form factors to the parametrization. Here, by considering the quark part of the symmetric EMT current we calculate the six transition gravitational form factors of the $N^* \rightarrow N$ by means of LCSR formalism \cite{Braun:1988qv, Balitsky:1989ry, Chernyak:1990ag} in full QCD. Using these form factors as the main input parameters, we  are able to extract the gravitational form factors of $N^*$ state. We find the fit functions of the form factors in terms of $Q^2$ and use them to estimate the pressure and  energy density at the center of $N^*$ state and calculate its mechanical radius.

Intensive spectroscopy programs are currently underway at different particle factories in the quest for undiscovered excited  baryons, especially the excited nucleons. Experiments  such as CEBAF at JLAB in the US, ELSA at Bonn University in Germany, and MAMI at the Johannes Gutenberg University at Mainz in Germany, focus on the s-channel excitation of nucleons to $N^*$ and $\Delta^*$  states \cite{Burkert:2019kxy}. In order to access  GFFs of  $N^* \rightarrow N $  experimentally, the nucleon to  $N^*$ transition generalized parton distributions have to be studied \cite{Burkert:2019kxy}. By this the mechanical properties of excited nucleons will be accessible, as well. 

The rest of the paper is organized as follows. In next section,  we calculate the  LCSRs for the six  transition
GFFs for  $N^* \rightarrow N $.  Using the obtained results, we also calculate LCSRs for GFFs of $ N^{*}$ state in  the same section. In  section \ref{secIII}, we numerically analyze the  obtained sum rules to find  the  GFFs  in terms of $ Q^2 $  using the DAs for  $N^*$ state.  Last section encompasses our  concluding notes.

\section{Formalism} \label{secII}
In this section, first we revisit the transitional gravitational FFs of the $ N^{*} \rightarrow N$  studied  in Ref. \cite{Ozdem:2019pkg} then obtain the LCSR for the GFFs of the $ N^{*}$ state.

 \subsection{Revisiting the  transition \texorpdfstring{$ N^{*} \rightarrow N$}{} GFFs} 
 The starting point is to consider an appropriate  LCSR correlation function,  a time ordering product of the nucleon and energy-momentum tensor currents sandwiched between the vacuum and $N^*$ states: 
 \begin{equation}\label{corf}
	\Pi_{\mu\nu}(p,q)=i\int d^4 x e^{iqx} \langle 0 |\mathcal{T}[J_{N}(0)T_{\mu \nu}^q(x)]|N^*(p)\rangle,
\end{equation}
where 
 $J_{N}(0)$ is  the nucleon's interpolating  current and   the $T_{\mu\nu}^q(x)$ is the quark part of the EMT current. Note that the EMT current has a gluonic part  as well (for instance see \cite{Azizi:2019ytx} ). Taking  the gluonic part of EMT current requires knowledge of the quark-gluon mixed DAs of the
$N^* $, which unfortunately are not available.  Considering the gluonic part leads to  five (or more)-particle distribution amplitudes,   contributions of which  are expected to be small  (see for instance \cite{Diehl:1998kh,Braun:2006hz}). Hence, we ignore from the gluonic part of the EMT current in the present study.

Performing the four-integral over $x$  the correlation function takes the form

 \begin{align}\label{phys}
 \Pi_{\mu\nu}^{Had}(p,q) &=\sum_{s{'}} \frac{\langle0|J_N|{N(p',s')}\rangle\langle {N(p',s')}
 |T_{\mu \nu}^q|N^*(p,s)\rangle}{m^2_{N}-p'^2} 
 +...,
\end{align}
where dots represent the contributions of the higher states and continuum and we set the final threshold to include only the nucleon in the final state. To proceed, we need to introduce the residue of the nucleon ($ \lambda_N $):
 \begin{align}
 \langle0|J_N|{N(p',s')}\rangle &= \lambda_N u_N(p',s'),
  \end{align}
   where,  $ u_N(p',s') $   is the Dirac spinor of the momentum $p' $   and spin $s' $. To go further, we introduce the  matrix element of the quark part of the EMT current. As the quark part of the current  alone is not conserved, this matrix element is parameterized in terms of six form factors by imposing the conditions of Lorentz invariance, discrete space-time symmetries and the equations of motion (Gordon identities) \cite{Polyakov:2020rzq}: 
\begin{widetext}
 \begin{align}\label{GFFs}
 \langle N(p^\prime,s')|T^{\mu\nu}_q|N^*(p,s)\rangle &=
 \bar{u}_N(p^\prime,s')\bigg[\frac{F_1(Q^2)}{\bar m^3}\bigg\{\Delta^2\, \tilde P_{\{\mu}\tilde P_{\nu\}} 
 - (m_{N^*}^2 - m_N^2)  \Delta_{\{\mu}\Delta_{\nu\}} + \frac{(m_{N^*}^2 - m_N^2)^2}{4} g_{\mu\nu}   \bigg\}\nonumber\\
 &+\frac{F_2(Q^2)}{\bar m^2}\bigg\{ \Delta^2  \gamma_{\{\mu}\tilde P_{\nu\}} - (m_{N^*}+m_N) \Delta_{\{\mu}\tilde P_{\nu\}}
 -  \frac{(m_{N^*}^2 - m_N^2)}{2}\bigg(  \gamma_{\{\mu}\Delta_{\nu\}} \nonumber\\
 &-  (m_{N^*}+m_N) g_{\mu\nu} \bigg) \bigg\}
 +\frac{F_3(Q^2)}{\bar m}\bigg\{ \Delta_{\mu}\Delta_{\nu}- \Delta^2  g_{\mu\nu} \bigg\}
 \nonumber\\
  &+ \bar m \bar C_1  (Q^2)  g_{\mu\nu} +\bar C_2  (Q^2) \gamma_{\{\mu}\tilde P_{\nu\}} 
  +\bar C_3 (Q^2)  \gamma_{\{\mu}\Delta_{\nu\}} \bigg] \gamma_5 u_{N^*}(p,s),
 \end{align}
 
\end{widetext}
where $m_{N^*}$ and   $m_N$  are the mass of the initial and final nucleon states, respectively;  and  $\bar m = {(m_{N^*}+ m_N)/2}$, 
$\tilde P= (p'+p)/2$, 
$\Delta = p'-p$,  
$X_{\{\mu}Y_{\nu\}} = (X_{\mu}Y_{\nu}+ X_{\nu}Y_{\mu})/2$ and  $Q^2=- \Delta^2$.
Here, $F_1(Q^2)  $, $F_2(Q^2)  $, $F_3(Q^2) $,  $\bar C_1  (Q^2) $, $\bar C_2  (Q^2) $ and $\bar C_3 (Q^2) $ are the transition GFFs. Note that by introducing  $\bar m $ into the above definition at different places we tried to make the form factors dimensionless, which is not done in  \cite{Polyakov:2020rzq}.

 For further simplification, we introduce the Dirac summation  over spin of the nucleon:
\begin{eqnarray}\label{sspin}
     \sum _{s'} u_N(p',s')\, \bar u_N(p',s') &=& \pslash'+m_N.
\end{eqnarray}

Using  Eqs. \eqref{GFFs} and \eqref{sspin} in Eq. \eqref{phys}, we acquire the hadronic representation of the correlation function in terms of FFs and other  hadronic parameters as
\begin{widetext}

 \begin{align}\label{physson}
\Pi_{\mu\nu}^{Had} (p,q)&= \frac{\lambda_N}{m^2_{N}-p'^2}( \pslash'+m_N)\bigg[\frac{F_1(Q^2)}{\bar m^3}\bigg\{\Delta^2\, \tilde P_{\{\mu}\tilde P_{\nu\}} 
 - (m_{N^*}^2 - m_N^2)  \Delta_{\{\mu}\Delta_{\nu\}} + \frac{(m_{N^*}^2 - m_N^2)^2}{4} g_{\mu\nu}   \bigg\}\nonumber\\
 &+\frac{F_2(Q^2)}{\bar m^2}\bigg\{ \Delta^2  \gamma_{\{\mu}\tilde P_{\nu\}} - (m_{N^*}+m_N) \Delta_{\{\mu}\tilde P_{\nu\}}
 -  \frac{(m_{N^*}^2 - m_N^2)}{2}\bigg(  \gamma_{\{\mu}\Delta_{\nu\}} \nonumber\\
 &-  (m_{N^*}+m_N) g_{\mu\nu} \bigg) \bigg\}
 +\frac{F_3(Q^2)}{\bar m}\bigg\{ \Delta_{\mu}\Delta_{\nu}- \Delta^2  g_{\mu\nu} \bigg\}
 \nonumber\\
  &+ \bar m \bar C_1  (Q^2)  g_{\mu\nu} +\bar C_2  (Q^2) \gamma_{\{\mu}\tilde P_{\nu\}} 
  +\bar C_3 (Q^2)  \gamma_{\{\mu}\Delta_{\nu\}} \bigg] \gamma_5 u_{N^*}(p,s)+.....
\end{align}

\end{widetext}
One can write the hadronic side of the correlation function in terms of  different Lorentz structures:
\begin{align}
 \Pi_{\mu\nu}^{Had} (p,q) &=\Pi_1^{Had}(Q^2)\,(  p^{\prime}_\mu q_\nu \qslash \gamma_5 + p^{\prime}_\nu q_\mu \qslash \gamma_5 ) \nonumber\\
 & + \Pi_2^{Had}(Q^2)\,(  p^{\prime}_\mu \gamma_\nu \qslash \gamma_5 + p^{\prime}_\nu \gamma_\mu \qslash \gamma_5 ) \nonumber\\
 &+ \Pi_3^{Had}(Q^2)\, q_\mu q_\nu \qslash \gamma_5 \nonumber\\
 &+ \Pi_4^{Had}(Q^2)\,g_{\mu\nu}\qslash \gamma_5\nonumber\\
 &+ \Pi_5^{Had}(Q^2)\, p^{\prime}_\mu p^{\prime}_\nu  \gamma_5 \nonumber\\
 &+ \Pi_6^{Had}(Q^2)\,( q_\mu \gamma_\nu \qslash \gamma_5 +  q_\nu \gamma_\mu \qslash \gamma_5 )\nonumber\\
& +....,
\end{align}
where the invariant functions  $ \Pi_i^{Had}(Q^2) $ are functions of GFFs.

The next step is to evaluate the correlation function in terms of QCD fundamental parameters. To this end, we replace the currents in correlation function with their explicit expressions in terms of quark fields, which are given as
\begin{align}
\label{intpol}
 J_N(0) &= 2\,\epsilon^{abc}\bigg[\big[u^{aT}(0) C  d^b(0)\big]\gamma_5 u^c(0) 
 + t\,\big[u^{aT}(0) C \gamma_5  d^b(0)\big] u^c(0)\bigg],\nonumber\\
 T_{\mu\nu}^q (x) &= \frac{i}{2}\bigg[\bar{u}^d(x)\overleftrightarrow{D}_\mu(x) \gamma_\nu u^d(x) 
 + \bar{d}^e(x)\overleftrightarrow{D}_\mu(x) \gamma_\nu d^e(x) 
+(\mu \leftrightarrow \nu) \bigg],
\end{align}
where $C$ is the charge conjugation operator, $t$ is the arbitrary mixing parameter; and $a$, $b$, $c$, $d$, $e$ are color indices.
The two-sides covariant derivative is expressed as
\begin{align}
 \overleftrightarrow{D}_\mu(x) &=\frac{1}{2} \Big[ \overrightarrow{D}_\mu(x) - \overleftarrow{D}_\mu(x) \Big],
\end{align}
with 
\begin{align}
 \overrightarrow{D}_\mu(x) &= \overrightarrow{\partial}_\mu(x)+igA_\mu(x), \\
\overleftarrow{D}_\mu(x) &= \overleftarrow{\partial}_\mu(x) -igA_\mu(x), 
\end{align}
 where $A_\mu$ is the gluon field. As we also previously mentioned, the gluon field  is neglected as the  quark-gluon mixed DAs of the $N^*$ state are unknown.
Thus, we take  the quark part of the EMT current in Eq. (\ref{intpol}).
 By insertion of the explicit forms of the interpolating field into the correlation function and performing contractions among the quark fields using the Wick theorem, we obtain the QCD representation in terms of quark propagators, DAs of $N^*$ state and their derivatives. As a result we have
 
 \begin{widetext}
 
\begin{align}\label{corrfunc}
	\Pi_{\mu\nu}^{QCD}(p,q)&=-\int d^4 x e^{iqx}\Bigg[\bigg\{ (\gamma_5)_{\gamma\delta}\, C_{\alpha\beta}\, ( \overleftrightarrow{D}_\mu (x)\gamma_\nu)_{\omega \rho} 
	+
	t\, (I)_{\gamma\delta}\,(C \gamma_5)_{\alpha\beta}\,( \overleftrightarrow{D}_\mu(x) \gamma_\nu)_{\omega \rho}\nonumber\\
	&+
	(\gamma_5)_{\gamma\delta} \, C_{\alpha\beta} \, ( \overleftrightarrow{D}_\nu (x)\gamma_\mu)_{\omega \rho} 
    +
	t\,(I)_{\gamma\delta}\,(C \gamma_5)_{\alpha\beta}\, ( \overleftrightarrow{D}_\nu (x)\gamma_\mu)_{\omega \rho}
	\bigg\}\nonumber\\
   & \times	
      \bigg\{\Big (\delta_\sigma^\alpha \delta_\theta^\rho \delta_\phi^\beta S_u(-x)_{\delta \omega}
     +\, \delta_\sigma^\delta \delta_\theta^\rho \delta_\phi^\beta S_u(-x)_{\alpha \omega}\Big)
     \langle 0|\epsilon^{abc} u_{\sigma}^a(0) u_{\theta}^b(x) d_{\phi}^c(0)|N^*(p)\rangle \nonumber\\
     &+\delta_\sigma^\alpha \delta_\theta^\delta \delta_\phi^\rho S_d(-x)_{\beta \omega} 
    \,\langle 0|\epsilon^{abc} u_{\sigma}^a(0) u_{\theta}^b(0) d_{\phi}^c(x)|N^*(p)\rangle\bigg\}
   \Bigg],
\end{align}

\end{widetext}
where $S_q(x)$ is the light quark propagator given by
\begin{align}
\label{edmn09}
S_{q}(x) &= 
\frac{1}{2 \pi^2 x^2}\Big( i \frac{{\xslash}}{x^{2}}-\frac{m_{q}}{2 } \Big)
- \frac{\langle  \bar qq \rangle }{12} \Big(1-i\frac{m_{q} \xslash}{4}   \Big)
- \frac{\langle \bar q \sigma.G q \rangle }{192}x^2  \Big(1-i\frac{m_{q} \xslash}{6}   \Big)
-\frac {i g_s }{32 \pi^2 x^2} ~G^{\mu \nu} (x) \bigg[\rlap/{x}
\sigma_{\mu \nu} +  \sigma_{\mu \nu} \rlap/{x}
 \bigg].
\end{align}
We set  $m_q = 0$ and  the terms proportional  to $\langle  \bar qq \rangle$ and $\langle \bar q \sigma.G q \rangle$ are killed  after Borel transformation. Hence, only  the first term in  the light quark propagator gives contribution to  the calculations.
The matrix element  $\langle 0| \epsilon^{abc} u_{\sigma}^a(a_1 x) u_{\theta}^b(a_2 x) d_{\phi}^c(a_3 x)|N^*(p)\rangle$ 
 is written in terms of  $N^*$ DAs, i.e., 
\bea\label{wave func}
&& 4 \langle 0 \vel \epsilon^{abc} u_\alpha^a(a_1 x) d_\beta^b(a_2 x)
d_\gamma^c(a_3 x) \ver N^\ast(p)\rangle\nnb\\
\es \mathcal{S}_1 m_{N^\ast}C_{\alpha\beta}N_{\gamma}^\ast -
\mathcal{S}_2 m_{N^\ast}^2 C_{\alpha\beta}(\rlap/x N^\ast)_{\gamma}\nnb\\
\ar \mathcal{P}_1 m_{N^\ast}(\gamma_5 C)_{\alpha\beta}(\gamma_5 N^\ast)_{\gamma} +
\mathcal{P}_2 m_{N^\ast}^2 (\gamma_5 C)_{\alpha\beta}(\gamma_5 \rlap/x N^\ast)_{\gamma} -
\left(\mathcal{V}_1 + \frac{x^2 m_{N^\ast}^2}{4} \mathcal{V}_1^M \right) 
(\rlap/p C)_{\alpha\beta} N_{\gamma}^\ast \nnb\\
\ar \mathcal{V}_2 m_{N^\ast}(\rlap/p C)_{\alpha\beta}(\rlap/x N^\ast)_{\gamma} + 
\mathcal{V}_3 m_{N^\ast}(\gamma_\mu C)_{\alpha\beta} (\gamma^\mu N^\ast)_{\gamma} -
\mathcal{V}_4 m_{N^\ast}^2 (\rlap/x C)_{\alpha\beta} N_{\gamma}^\ast \nnb\\
\ek \mathcal{V}_5 m_{N^\ast}^2(\gamma_\mu C)_{\alpha\beta}
(i \sigma^{\mu\nu} x_\nu N^\ast)_\gamma +
\mathcal{V}_6 m_{N^\ast}^3 (\rlap/x C)_{\alpha\beta}(\rlap/x N^\ast)_{\gamma} \nnb \\
\ek \left(\mathcal{A}_1 + \frac{x^2m_{N^\ast}^2}{4}\mathcal{A}_1^M\right)
(\rlap/p \gamma_5 C)_{\alpha\beta} (\gamma N^\ast)_{\gamma} +
\mathcal{A}_2 m_{N^\ast}(\rlap/p \gamma_5 C)_{\alpha\beta} (\rlap/x \gamma_5 N^\ast)_{\gamma} +
\mathcal{A}_3 m_{N^\ast}(\gamma_\mu\gamma_5 C)_{\alpha\beta}
(\gamma^\mu \gamma_5 N^\ast)_{\gamma}\nnb\\
\ek \mathcal{A}_4 m_{N^\ast}^2(\rlap/x \gamma_5 C)_{\alpha\beta}
(\gamma_5 N^\ast)_{\gamma} -
\mathcal{A}_5 m_{N^\ast}^2(\gamma_\mu \gamma_5 C)_{\alpha\beta}
(i \sigma^{\mu\nu} x_\nu \gamma_5 N^\ast)_{\gamma} +
\mathcal{A}_6 m_{N^\ast}^3(\rlap/x \gamma_5 C)_{\alpha\beta}
(\rlap/x \gamma_5 N^\ast)_{\gamma}\nnb\\
\ek \left(\mathcal{T}_1 + \frac{x^2m_{N^\ast}^2}{4}\mathcal{T}_1^M \right)
(i \sigma_{\mu\nu} p_\nu C)_{\alpha\beta} (\gamma^\mu N^\ast)_{\gamma} +
\mathcal{T}_2 m_{N^\ast} (i \sigma_{\mu\nu} x^\mu p^\nu C)_{\alpha\beta}
N_{\gamma}^\ast\nnb\\
\ar \mathcal{T}_3 m_{N^\ast}(\sigma_{\mu\nu} C)_{\alpha\beta}
(\sigma^{\mu\nu} N^\ast)_{\gamma} +
\mathcal{T}_4 m_{N^\ast} (\sigma_{\mu\nu} p^\nu C)_{\alpha\beta}
(\sigma^{\mu\rho} x_\rho N^\ast)_{\gamma} \nnb\\
\ek \mathcal{T}_5 m_{N^\ast}^2 (i\sigma_{\mu\nu} x^\nu C)_{\alpha\beta}
(\gamma^\mu N^\ast)_{\gamma} -  
\mathcal{T}_6 m_{N^\ast}^2 (i \sigma_{\mu\nu} x^\mu p^\nu C)_{\alpha\beta}
(\rlap/x N^\ast)_{\gamma}\nnb\\
\ek \mathcal{T}_7 m_{N^\ast}^2 (\sigma_{\mu\nu} C)_{\alpha\beta}
(\sigma^{\mu\nu} \rlap/x N^\ast)_{\gamma} +
\mathcal{T}_8 m_{N^\ast}^3 (\sigma_{\mu\nu} x^\nu C)_{\alpha\beta}
(\sigma^{\mu\rho} x_\rho N^\ast)_{\gamma}~.\nnb
\eea 
 The calligraphic functions in the above matrix element  are expressed in terms of  the functions  having  definite twists as
  \begin{flalign}
		   \mathcal{S}_1 =& S_1,\hspace{3.5cm}                    2p.x\mathcal{S}_2=S_1-S_2,\nonumber\\
		   \mathcal{P}_1=&P_1, \hspace{3.5cm}                     2p.x\mathcal{P}_2=P_1-P_2,\\
           \mathcal{V}_1=&V_1,\hspace{3.5cm}                      2p.x\mathcal{V}_2=V_1-V_2-V_3, \nonumber\\
           2\mathcal{V}_3=&V_3,\hspace{3.5cm}                     4p.x\mathcal{V}_4=-2V_1+V_3+V_4+2V_5,\nonumber\\
           4p.x\mathcal{V}_5=&V_4-V_3,\hspace{2.5cm}              4(p.x)^2\mathcal{V}_6=-V_1+V_2+V_3+V_4 + V_5-V_6\\
           \mathcal{A}_1=&A_1, \hspace{3.5cm}                     2p.x\mathcal{A}_2=-A_1+A_2-A_3,\nonumber\\
        2\mathcal{A}_3=&A_3,\hspace{3.5cm}                        4p.x\mathcal{A}_4=-2A_1-A_3-A_4+2A_5, \nonumber\\
        4p.x\mathcal{A}_5=&A_3-A_4, \hspace{2.5cm}                4(p.x)^2\mathcal{A}_6=A_1-A_2+A_3+A_4-A_5+A_6\\
            \mathcal{T}_1=&T_1, \hspace{3.5cm}                    2p.x\mathcal{T}_2=T_1+T_2-2T_3, \nonumber\\
            2\mathcal{T}_3=&T_7,\hspace{3.5cm}                    2p.x\mathcal{T}_4=T_1-T_2-2T_7,\nonumber\\
        2p.x\mathcal{T}_5=&-T_1+T_5+2T_8,\hspace{1cm}             4(p.x)^2\mathcal{T}_6=2T_2-2T_3-2T_4+2T_5+2T_7+2T_8,        \nonumber\\ 
        4p.x \mathcal{T}_7=&T_7-T_8,\hspace{2.5cm}                4(p.x)^2\mathcal{T}_8=-T_1+T_2 +T_5-T_6+2T_7+2T_8,
\end{flalign}
where $V_i, A_i, T_i,S_i $ and $P_i$  are vector, axialvector, tensor, scalar and pesudoscalar  DAs of  $N^*$  state, respectively. Their explicit forms together with all the relative parameters are given in  Ref \cite{maxiphd}.

Making use of the  DAs of $N^*$ state and  performing the  Fourier transformations, the QCD side of the correlation function is obtained, which reads 
\begin{align}
 \Pi_{\mu\nu}^{QCD} (p,q) &= \Pi_1^{QCD}(Q^2)\,(  p^{\prime}_\mu q_\nu \qslash \gamma_5 + p^{\prime}_\nu q_\mu \qslash \gamma_5 ) \nonumber\\
 & + \Pi_2^{QCD}(Q^2)\,(  p^{\prime}_\mu \gamma_\nu \qslash \gamma_5 + p^{\prime}_\nu \gamma_\mu \qslash \gamma_5 ) \nonumber\\
 &+ \Pi_3^{QCD}(Q^2)\, q_\mu q_\nu \qslash \gamma_5 \nonumber\\
 &+ \Pi_4^{QCD}(Q^2)\,g_{\mu\nu}\qslash \gamma_5\nonumber\\
 &+ \Pi_5^{QCD}(Q^2)\, p^{\prime}_\mu p^{\prime}_\nu  \gamma_5 \nonumber\\
 &+ \Pi_6^{QCD}(Q^2)\,( q_\mu \gamma_\nu \qslash \gamma_5 +  q_\nu \gamma_\mu \qslash \gamma_5 )\nonumber\\
& +....
\end{align}

The desired  LCSR for the $N^* \rightarrow N $ transition GFFs are achieved by equating the coefficients of 
different Lorentz structures from both the hadronic and QCD sides of the correlation function.  To suppress the contributions of the higher states and continuum, Borel transformation and continuum subtraction are applied.
We shall note that we use the structures $(  p^{\prime}_\mu q_\nu \qslash \gamma_5 + p^{\prime}_\nu q_\mu \qslash \gamma_5 )$, $(  p^{\prime}_\mu \gamma_\nu \qslash \gamma_5 + p^{\prime}_\nu \gamma_\mu \qslash \gamma_5 )$, 
$q_\mu q_\nu \qslash \gamma_5$,  $g_{\mu \nu} \qslash\gamma_5$, $ p^{\prime}_\mu p^{\prime}_\nu  \gamma_5$ and  $( q_\mu \gamma_\nu \qslash \gamma_5 +  q_\nu \gamma_\mu \qslash \gamma_5 )$
to find the LCSR for  the $ N^{*} \rightarrow N$ transition GFFs, $F_1(Q^2)  $, $F_2(Q^2)  $, $F_3(Q^2) $,  $\bar C_1  (Q^2) $, $\bar C_2  (Q^2) $ and $\bar C_3 (Q^2) $, respectively.
 Hence, 

\begin{align}
 F_1(Q^2)   &=-\frac{\bar m^3 }{\lambda_N (m_N-m_{N^*})^2}  \,\,e^{\frac{m_N^2}{M^2}} \varPi_1^{QCD},\\  \label{F1FF}
F_2(Q^2)  &=-\frac{ \bar m^2}{\lambda_N } \,e^{\frac{m_N^2}{M^2}} \varPi_2^{QCD},\\  \label{F2FF}
 F_3(Q^2)  &=\frac{\bar m}{\lambda_N }\,e^{\frac{m_N^2}{M^2}}\, \varPi_3^{QCD}, \\ \label{F3FF}
 \bar C_1 (Q^2)  &= - \frac{1}{\bar m \,\lambda_N}\,e^{\frac{m_N^2}{M^2}}\, \varPi_4^{QCD},\\  
 \bar C_2 (Q^2)  &=  \frac{1}{2\,\lambda_N}\,e^{\frac{m_N^2}{M^2}}\, \varPi_5^{QCD},\\      
 \label{C3FF}  \bar C_3 (Q^2)  &= -\frac{1}{\lambda_N} \,e^{\frac{m_N^2}{M^2}}\,  \varPi_6^{QCD}. 
\end{align}

The $\Pi_i^{QCD}$ functions  are quite lengthy, therefore the explicit expressions of the these functions are not presented here.  The  Borel and subtraction procedures are applied  using the following relations (see for instance Ref. \cite{Braun:2006hz}):

\begin{align}
		\int dx \frac{\rho(x)}{(q-xp)^2}&\rightarrow -\int_{x_0}^1\frac{dx}{x}\rho(x) e^{-s(x)/M^2}, \nonumber		\\
		\int dx \frac{\rho(x)}{(q-xp)^4}&\rightarrow \frac{1}{M^2} \int_{x_0}^1\frac{dx}{x^2}\rho(x) e^{-s(x)/M^2}
		+\frac{\rho(x_0)}{Q^2+x_0^2 m_{N^{*}}^2} e^{-s_0/M^2},\nonumber\\
	\int dx \frac{\rho(x)}{(q-xp)^6}&\rightarrow -\frac{1}{2M^4}\int_{x_0}^1\frac{dx}{x^3}\rho(x) e^{-s(x)/M^2}
		-\frac{1}{2M^2}\frac{\rho(x_0)}{x_0(Q^2+x_0^2m_{N^{*}}^2)}e^{-s_0/M^2}\nonumber\\
		&+\frac{1}{2}\frac{x_0^2}{Q^2+x_0^2m_{N^{*}}^2}\bigg[\frac{d}{dx_0}\frac{\rho(x_0)}{x_0(Q^2+x_0^2m_{N^{*}}^2)}\bigg]e^{-s_0/M^2}
	\label{subtract3}
\end{align}
Here
\begin{eqnarray}
s(x)=(1-x)m_{N^{*}}^2+\frac{1-x}{x}Q^2,
\end{eqnarray}
where $M^2$ is the Borel mass  parameter, which appears following the application of the Borel transformation with respect to the $p^{\prime 2}$, and $x_0$ is the solution of the equation  $s(x)=s_0$, i.e.,
\begin{eqnarray}
x_0&=&\Big[\sqrt{(Q^2+s_0-m_{N^{*}}^2)^2+4m_{N^{*}}^2 Q^2}-(Q^2+s_0-m_{N^{*}}^2)\Big]/2m_{N^{*}}^2,
\end{eqnarray}
with $s_0$ being the continuum threshold parameter.

\subsection{GFFs of the \texorpdfstring{$ N^{*}$ state}{}}

\begin{equation}\label{corf1}
	\Pi_{\mu\nu}^{N^*}(p,q)=i\int d^4 x e^{iqx} \langle 0 |\mathcal{T}[J_{N^*}(0)T_{\mu \nu}^q(x)]|N^*(p)\rangle,
\end{equation}
where, as we previously mentioned,  $ N^{*}$ state couples to the same as the nucleon, i.e.,
\begin{align}
\label{intpol111}
 J_{N^*}(0) &= 2\,\epsilon^{abc}\bigg[\big[u^{aT}(0) C  d^b(0)\big]\gamma_5 u^c(0) 
 + t\,\big[u^{aT}(0) C \gamma_5  d^b(0)\big] u^c(0)\bigg].
 \end{align}
 Using this three-particle interpolating current, the mass of the $N^*$ state was already  extracted to be well consistent with the experimental data  in Refs.  \cite{Jido:1996ia, Kondo:2005ur}, using the QCD sum rule method. A reasonable and compatible result  with those extracted from the experimental data for the strong coupling of  $N^*$ to  $N\pi$ was also obtained in Ref. \cite{Azizi:2015jya} using the same method.   Although  this state  could be the lowest $ L = 1$ orbital excited $ |uud \rangle$  state with a large
admixture of $|[ud][us]\bar s \rangle$  pentaquark component having $[ud]$,$ [us]$ and $\bar s$ in the ground state according to  Ref. \cite{Zou:2010tc}, we consider it as a three-particle state interpolating by the above current based on the results of   \cite{Jido:1996ia, Kondo:2005ur,Azizi:2015jya}.

 Now, we saturate the correlation function with the intermediate hadronic states, which receives contributions from both the $ N^{*}$ and $ N$ states when we set the final  thrashed  considering the   $ N^{*}$ state. As a result, we get
 \begin{align}\label{phys1}
 \Pi_{\mu\nu}^{Had-N^*}(p,q) &= \frac{\langle0|J_N|{N(p',s')}\rangle\langle {N(p',s')}
 |T_{\mu \nu}^q|N^*(p,s)\rangle}{m^2_{N}-p'^2} \nonumber\\
 & + \frac{\langle0|J_N|{N^*(p',s')}\rangle\langle {N^*(p',s')}
 |T_{\mu \nu}^q|N^*(p,s)\rangle}{m^2_{N^*}-p'^2} \nonumber\\
 &+....
\end{align}
This can be further simplified by introducing the matrix elements
\begin{align}\label{GFFs1}
 \langle0|J_N|{N^*(p',s')}\rangle &= \lambda_{N^*} \gamma_5 u_{N^*}(p',s'),\\
 \nonumber\\
  \langle N^*(p^\prime,s')|T_{\mu\nu}^q|N^*(p,s)\rangle &=
 \bar{u}_{N^*}(p^\prime,s')\bigg[A(Q^2)\frac{ \tilde P_\mu \tilde P_\nu}{m_{N^*}}
 +~i J(Q^2)\frac{(\tilde P_\mu \sigma_{\nu\rho}+\tilde P_\nu \sigma_{\mu\rho})\Delta^\rho}{2\,m_{N^*}}
    \nonumber\\
  & +~D(Q^2) \frac{\Delta_\mu \Delta_\nu- g_{\mu\nu} \Delta^2}{4\,m_{N^*}}
  +~ \bar c (Q^2) m_{N^*} g_{\mu \nu} \bigg]  u_{N^*}(p,s), \label{GFFson1}
 \end{align}
 where $  \lambda_{N^*}$  is the residue of  the $ N^{*}$ state; and  $ A(Q^2)$,  $ J(Q^2)$, $ D(Q^2)$  and  $ \bar c (Q^2)$  are its gravitational form factors. Using these matrix elements and those of the  $ N^{*}\rightarrow N$ transition from the previous section, we find
  \begin{align}\label{physson1}
\Pi_{\mu\nu}^{Had-N^*} (p,q)&= \frac{\lambda_N}{m^2_{N}-p'^2}( \pslash'+m_N)\bigg[\frac{F_1(Q^2)}{\bar m^3}\bigg\{\Delta^2\, \tilde P_{\{\mu}\tilde P_{\nu\}} 
 - (m_{N^*}^2 - m_N^2)  \Delta_{\{\mu}\Delta_{\nu\}} + \frac{(m_{N^*}^2 - m_N^2)^2}{4} g_{\mu\nu}   \bigg\}\nonumber\\
 &+\frac{F_2(Q^2)}{\bar m^2}\bigg\{ \Delta^2  \gamma_{\{\mu}\tilde P_{\nu\}} - (m_{N^*}+m_N) \Delta_{\{\mu}\tilde P_{\nu\}}
 -  \frac{(m_{N^*}^2 - m_N^2)}{2}\bigg(  \gamma_{\{\mu}\Delta_{\nu\}} \nonumber\\
 &-  (m_{N^*}+m_N) g_{\mu\nu} \bigg) \bigg\}
  +\frac{F_3(Q^2)}{\bar m}\bigg\{ \Delta_{\mu}\Delta_{\nu}- \Delta^2  g_{\mu\nu} \bigg\}
 \nonumber\\
  &+ \bar m \bar C_1(Q^2)  g_{\mu\nu} +\bar C_2(Q^2) \gamma_{\{\mu}\tilde P_{\nu\}} 
  +\bar C_3(Q^2)  \gamma_{\{\mu}\Delta_{\nu\}} \bigg] \gamma_5 u_{N^*}(p,s) \nonumber\\
  &+ \frac{\lambda_{N^*}}{m^2_{N^*}-p'^2} \gamma_5( \pslash'+m_{N^*})\bigg[A(Q^2)\frac{ \tilde P_\mu \tilde P_\nu}{m_{N^*}}
 +~i J(Q^2)\frac{(\tilde P_\mu \sigma_{\nu\rho}+\tilde P_\nu \sigma_{\mu\rho})\Delta^\rho}{2\,m_{N^*}}
    \nonumber\\
  & +~D(Q^2) \frac{\Delta_\mu \Delta_\nu- g_{\mu\nu} \Delta^2}{4\,m_{N^*}}
  +~ \bar c(Q^2) m_{N^*} g_{\mu \nu} \bigg]  u_{N^*}(p,s).
\end{align}

The above equation contains many structures, we use some of which to evaluate the GFFs of the  $ N^{*}$ state. We present the structures that are used explicitly and remove the others into the $...$ as follows

\begin{align}
 \Pi_{\mu\nu}^{Had-N^*} (p,q)&=\Pi_7^{Had}(Q^2)\, p^{\prime}_\mu p^{\prime}_\nu \qslash \gamma_5 + \Pi_8^{Had}(Q^2)\,p^{\prime}_\mu \gamma_\nu \gamma_5 \nonumber\\
& + \Pi_9^{Had}(Q^2)\,q_\mu q_\nu \qslash \gamma_5 + \Pi_{10}^{Had}(Q^2)\,g_{\mu\nu} \qslash\gamma_5 \nonumber\\
&+...,
\end{align}
where   the invariant functions $ \Pi_7^{Had}(Q^2)$,  $ \Pi_8^{Had}(Q^2)$,  $ \Pi_9^{Had}(Q^2)$ and $ \Pi_{10}^{Had}(Q^2)$ contain GFFs of both the  $ N^{*}-  N^{*}$  and $ N^{*}-  N$ transitions. 

 From a similar manner  the QCD side of the calculations is written in terms of the selected structures as 
\begin{align}
 \Pi_{\mu\nu}^{QCD-N^*} (p,q)&=\Pi_7^{QCD}(Q^2)\, p^{\prime}_\mu p^{\prime}_\nu \qslash \gamma_5 + \Pi_8^{QCD}(Q^2)\,p^{\prime}_\mu \gamma_\nu \gamma_5 \nonumber\\
& + \Pi_9^{QCD}(Q^2)\,q_\mu q_\nu \qslash \gamma_5 + \Pi_{10}^{QCD}(Q^2)\,g_{\mu\nu} \qslash\gamma_5 \nonumber\\
&+.... 
\end{align}
Matching the coefficients of the same structures from both sides, applying the Borel transformation with respect to $p^{\prime 2}$ and using the sum rules for the  $ N^{*}-  N$ transition from the previous section, we obtain the following sum rules for the GFFs of the $ N^{*}$ state:
\begin{align}
 A(Q^2) &=\frac{m_{N^*}}{\lambda_{N^*}}\,e^{\frac{m_{N^*}^2}{M^{*2}}}\, \varPi_7^{QCD}-\frac{\bar m^3}{\lambda_{N}(m_N-m_{N^*})^2} \,e^{\frac{m_{N}^2}{M^2}}\,\varPi_1^{QCD},\\
 \nonumber\\
 J(Q^2)&=\frac{m_{N^*}}{\lambda_{N^*}}\,e^{\frac{m_{N^*}^2}{M^{*2}}}\, \varPi_8^{QCD}-\frac{\bar m^2}{\lambda_{N}} \,e^{\frac{m_{N}^2}{M^2}}\,\varPi_2^{QCD}\\
 \nonumber\\
 D(Q^2)&= \frac{4 m_{N^*}}{\lambda_{N^*}}\,e^{\frac{m_{N^*}^2}{M^{*2}}}\, \varPi_9^{QCD}-\frac{m_{N^*}}{\lambda_{N^*}}\,e^{\frac{m_{N^*}^2}{M^{*2}}}\, \varPi_7^{QCD}-\frac{m_{N^*}}{\lambda_{N^*}}\,e^{\frac{m_{N^*}^2}{M^{*2}}}\, \varPi_8^{QCD}-
 \frac{\bar m}{\lambda_N }\,e^{\frac{m_N^2}{M^{*2}}}\, \varPi_3^{QCD}, \\
 \nonumber\\
 \bar c(Q^2)&=\frac{1}{\lambda_{N^*}\,m_{N^*}}\,e^{\frac{m_{N^*}^2}{M^{*2}}}\, \varPi_{10}^{QCD}- \frac{1}{\bar m \,\lambda_N}\,e^{\frac{m_N^2}{M^2}}\, \varPi_4^{QCD},
\end{align}
where  $M^{*2}$ is the new Borel parameter in $ N^{*}$ channel. The expressions in QCD sides also contain the new continuum threshold parameter $ s_0^{*}$, which will be fixed together with other auxiliary parameters in next section.

\section{Numerical Results}\label{secIII}
The sum rules for GFFs contain many input parameters that we need their numerical values. 
The DAs of  $N^*$ state and all the corresponding parameters are used from  Ref. \cite{maxiphd}. 
 We  use $m_N = 0.94$ GeV and $m_{N^*} = 1.51 \pm 0.01$ GeV~\cite{Tanabashi:2018oca} for the baryon masses.   
 For numerical evaluations of LCSR for GFFs of  $N^* \rightarrow N$  and $N^*-N^*$, we need the values of  the residues, $\lambda_N$ and $\lambda_{N^*}$ as well.  They are borrowed from Refs. \cite{Aliev:2011ku, Azizi:2015fqa}, which were determined from mass sum rules.

The LCSRs for GFFs under study are also functions of three auxiliary parameters that are included at different stages of the calculations: the mixing parameter $t$, the continuum threshold $s_0$ and the 
Borel mass parameter $M^2$. These are helping parameters that the physical observables should not depend on their variations. But in practice, the results show some dependencies on these objects. We need to search for their working intervals such that their variations in those windows lead to relatively small changes in the results. The weak dependence of the results on these parameters constitute the main part of the uncertainties in the results. The working intervals for these auxiliary parameters are determined by imposing the standard criteria of the method, namely, weak dependence of the results on these parameters, pole dominance and convergence of the series of the operator product expansion. The standard procedures lead to the intervals -0.31 $\leq cos\theta \leq$ -0.45 (where $t$= tan$\theta$ ),  $1.5$ GeV$^ 2$ $\leq s_0 \leq 2.0$ GeV$^ 2 $,  $2.3$ GeV$^ 2$ $\leq s^*_0 \leq 2.7$ GeV$^ 2 $,  $1.0$ GeV$^ 2$ $\leq M^2 \leq 2.0$ GeV$^ 2 $ and  $2.0$ GeV$^ 2$ $\leq M^{*2} \leq 3.0$ GeV$^ 2 $ for the auxiliary parameters. 


Having determined the working intervals for the helping parameters, we proceed to discuss the behavior of the form factors with respect to  $Q^2$. As usual, the LCSRs fo GFFs give reliable results at large $Q^2$ but they do not lead to safe results at lower values of $Q^2$. We need the values of GFFs at static limit,  $Q^2=0$. To this end, we use the beauty of the mathematics: we employ some fit functions to extrapolate the results to lower values of $Q^2$ such that the fits reproduce the LCSR results for  $Q^2 > 2$ GeV$^ 2$. Our numerical calculations depict that the GFFs of the $N^* \rightarrow N$ and $N^*-N^*$ transitions can properly be parameterized via the following multipole fit function:
\begin{align}
{\cal F}(Q^2)= \frac{{\cal F}(0)}{\Big(1+ e Q^2\Big)^p},
\end{align} 
where the fit parameters together with the values of the GFFs at $Q^2=0$ are given in tables \ref{fit_table}  and \ref{fit_table1}. Note that, we have only one result on the $D$-term of   $N^*$ available from the bag model~\cite{Neubelt:2019sou}. In this study, $D_{N^*}=-12.97$ is found, which is consistent with our prediction, $D(0)=-14.50 \pm 2.50$, within the errors.
 The presented errors in the tables are due to the uncertainties in the calculations of the working windows for the auxiliary parameters as well as those related to the uncertainties in the parameters of DAs and other inputs. The behavior of GFFs for $N^* \rightarrow N$ and and $N^*-N^*$ transitions  with respect to $Q^2$ are depicted in figures 1 and 2. These behaviors may be checked by different non-perturbative methods as well as in future experiments. As we previously mentioned, we have considered only the quark part of the EMT current, which is not conserved. The order of violation is determined from the value of $\bar c$ form factor.   The value  $\bar c(0) = -0.34 \pm 0.06$ obtained in the present study, shows that the violation in the conservation of the  quark part of the EMT current is considerably high in $N^*$ channel compared to that of the ground state nucleon obtained in Ref.  \cite{Azizi:2019ytx} but comparable with the prediction of, for instance, Ref. \cite{Neubelt:2019sou}. 

\begin{table}[htp]
\caption{Values for the parameters of the $N^*-N$ transition GFFs obtained  by applying the multipole fit functions.
}
\hspace*{-0.5cm}
\begin{tabular}{ |l|c|c|c|c|c|c|}
\hline\hline

Form Factors & ${\cal F}(0)$  & $e$(GeV$^{-2}$) & p  \\ \hline\hline
        $F_1$($Q^2$)      & $~~1.25 \pm 0.32$ &$1.08 \pm 0.09$ &$3.6 - 4.0$ \\
        $F_2$($Q^2$)     & $~~0.29 \pm 0.05$ & $0.90 \pm 0.07$ &$3.4 - 3.8$ \\
  $F_3$($Q^2$)   &  $-3.65 \pm 0.51$ &$ 0.94 \pm 0.05$ &$3.0-3.4$ 
   \\
  $\bar C_1$($Q^2$)  &$ -1.80 \pm 0.32$ &$0.95 \pm 0.07$ &$ 2.8-3.1$
       \\
  $\bar C_2$($Q^2$)  &$-0.30 \pm 0.06$ &$1.06 \pm 0.05$ &$2.7-3.0$
            \\
   $\bar C_3$($Q^2$)  &$-0.40 \pm 0.15$ &$ 1.20 \pm 0.10$ &$ 2.8-3.2$\\
\hline \hline
\end{tabular}
\label{fit_table}
\end{table}

\begin{table}[htp]
\caption{ Values for the parameters of the $N^*-N^*$ transition GFFs obtained  by applying the multipole fit functions.
}
\hspace*{-0.5cm}
\begin{tabular}{ |l|c|c|c|c|c|c|}
\hline\hline

Form Factors & ${\cal F}(0)$  & $e$(GeV$^{-2}$)  & p  \\ \hline\hline
        $A$($Q^2$)      & $~~0.72 \pm 0.10$ &$1.16 \pm 0.12$ &$3.3 - 3.7$ \\
$J$($Q^2$)     & $~~0.35 \pm 0.05$ & $1.10 \pm 0.10$ &       $3.2 - 3.6$ \\
  $D$($Q^2$)   &  $-14.50 \pm 2.50$ &$ 1.00 \pm 0.06$ &$3.6-4.0$ 
   \\
  $\bar c$($Q^2$)  &$ -0.34 \pm 0.06$ &$0.94 \pm 0.08$ &$ 3.6-4.0$
       \\
\hline \hline
\end{tabular}
\label{fit_table1}
\end{table}

After obtaining the GFFs of the $N^*-N^*$ transition, we can use them to compute some mechanical properties of  the $N^*$ state such as mechanical radius square ($\langle r^2_{\text{mech}}\rangle$) as well as energy ($\cal E$) and pressure ($p_0$) distributions at the center of the particle. The related formulas are given as \cite{Polyakov:2018zvc},
\begin{eqnarray}
\label{Pres}
&&p_0  =-\frac{1}{24\,\pi^2\, m_{N^*}} \int^{\infty}_{0} dy \,y\,\sqrt{y}\, [ D(y)-\bar c (y) ],\\
&&{\cal E}=\frac{m_{N^*}}{4\,\pi^2} \int^{\infty}_{0} dy \,\sqrt{y}\,\Big[ A(y) + \frac{y}{4m^2_H} [A(y) -2J(y)+ D(y)+ \bar c (y)]\Big],\\
&& \langle r^2_{\text{mech}}\rangle = \frac {6\, D(0)}{\int^{\infty}_{0} dy\, D(y)},
\end{eqnarray}
where $y = Q^2$.

\begin{table}[htp]
\addtolength{\tabcolsep}{2pt}
\begin{center}
\begin{tabular}{c|c|c|cccc}
\hline\hline
Transition &  ~~~ $p_0$  (GeV/fm$^3$)~~~&~~~ ${\cal E}$ (GeV/fm$^3$) ~~~&~~~$\langle {r^2_{\text{mech}}\rangle}$ (fm$^{2}$)~~~& \\[0.5ex]
\hline\hline

$N^*-N^*$ &$  1.36 \pm 0.33$ &$ 1.40 \pm 0.40$ &$0.65\pm 0.09$&  \\
\hline\hline
\end{tabular}
\caption{ The values of mechanical quantities for $N^*-N^*$ transition.}
\label{mech_table}
\end{center}
\end{table}
The numerical values for the mechanical quantities of the $N^*$ state are depicted in table \ref{mech_table}. From this table we see that the value of $\langle {r^2_{\text{mech}}\rangle}$ in $N^*$ channel is about  $20\%$ larger from that of the nucleon obtained in Ref.  \cite{Azizi:2019ytx}. Our results for all the mechanical properties may be checked via different approaches. 

 \begin{figure}[htp]
\centering
 \subfloat[]{\includegraphics[width=0.4\textwidth]{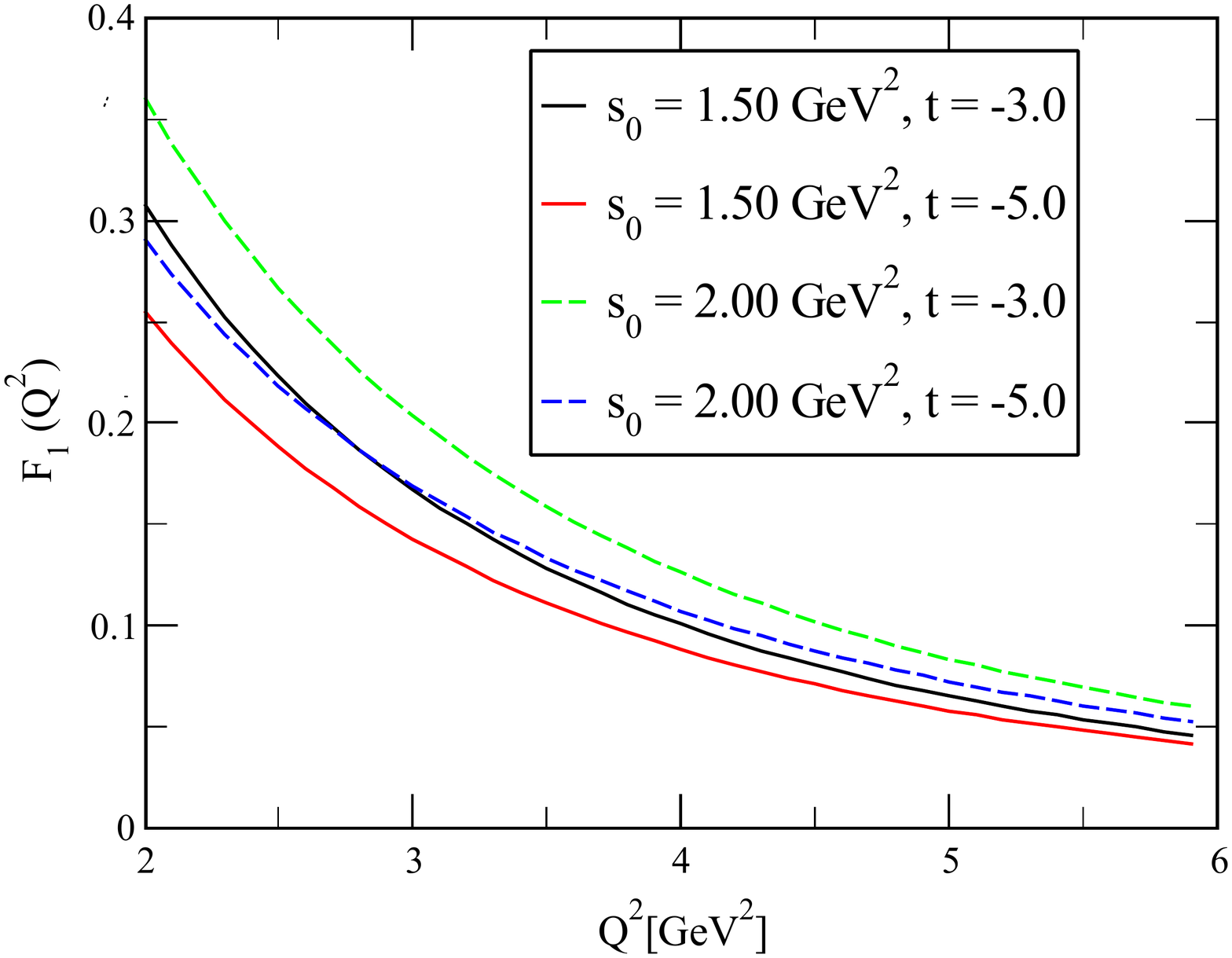}}~~~~
    \subfloat[]{\includegraphics[width=0.4\textwidth]{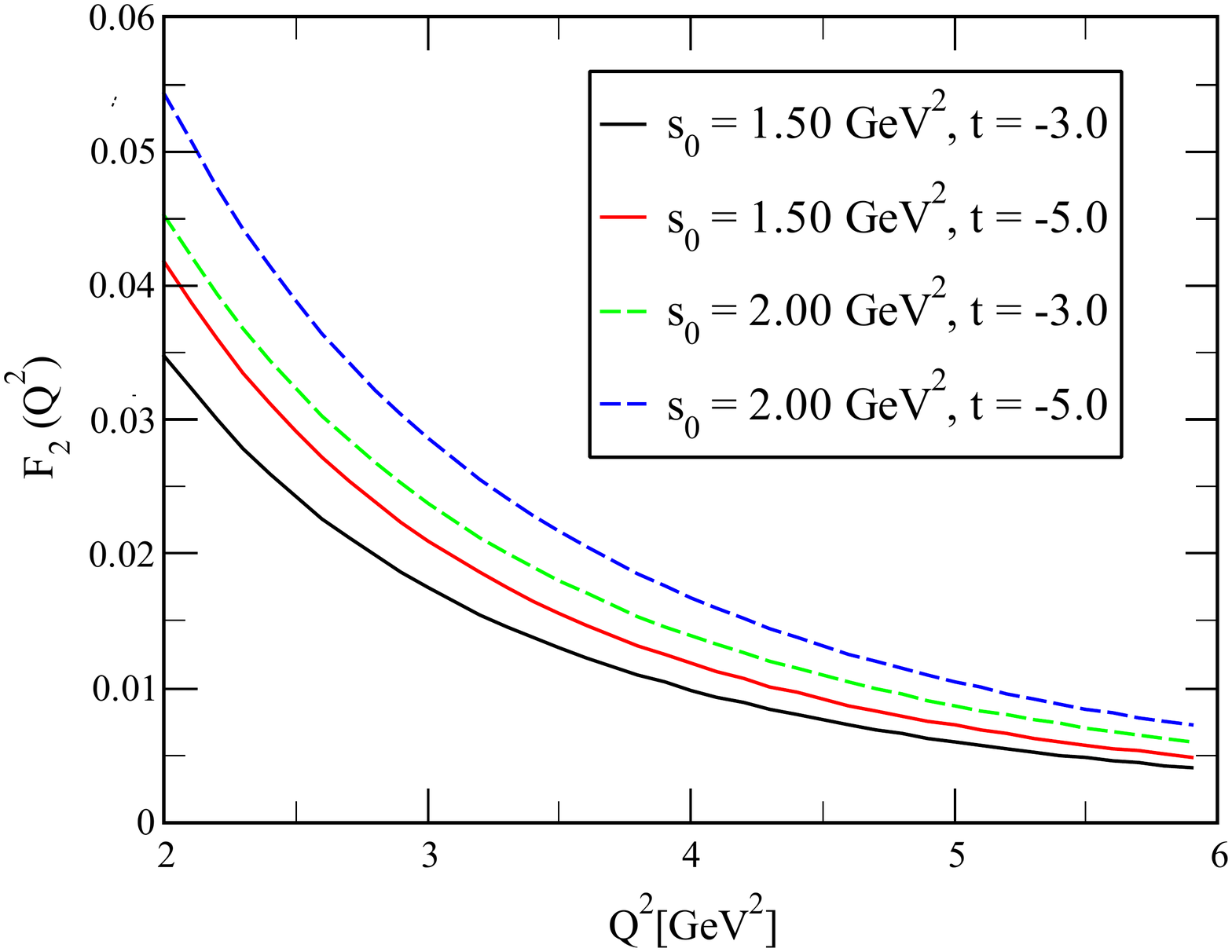}}\\
    \subfloat[]{\includegraphics[width=0.4\textwidth]{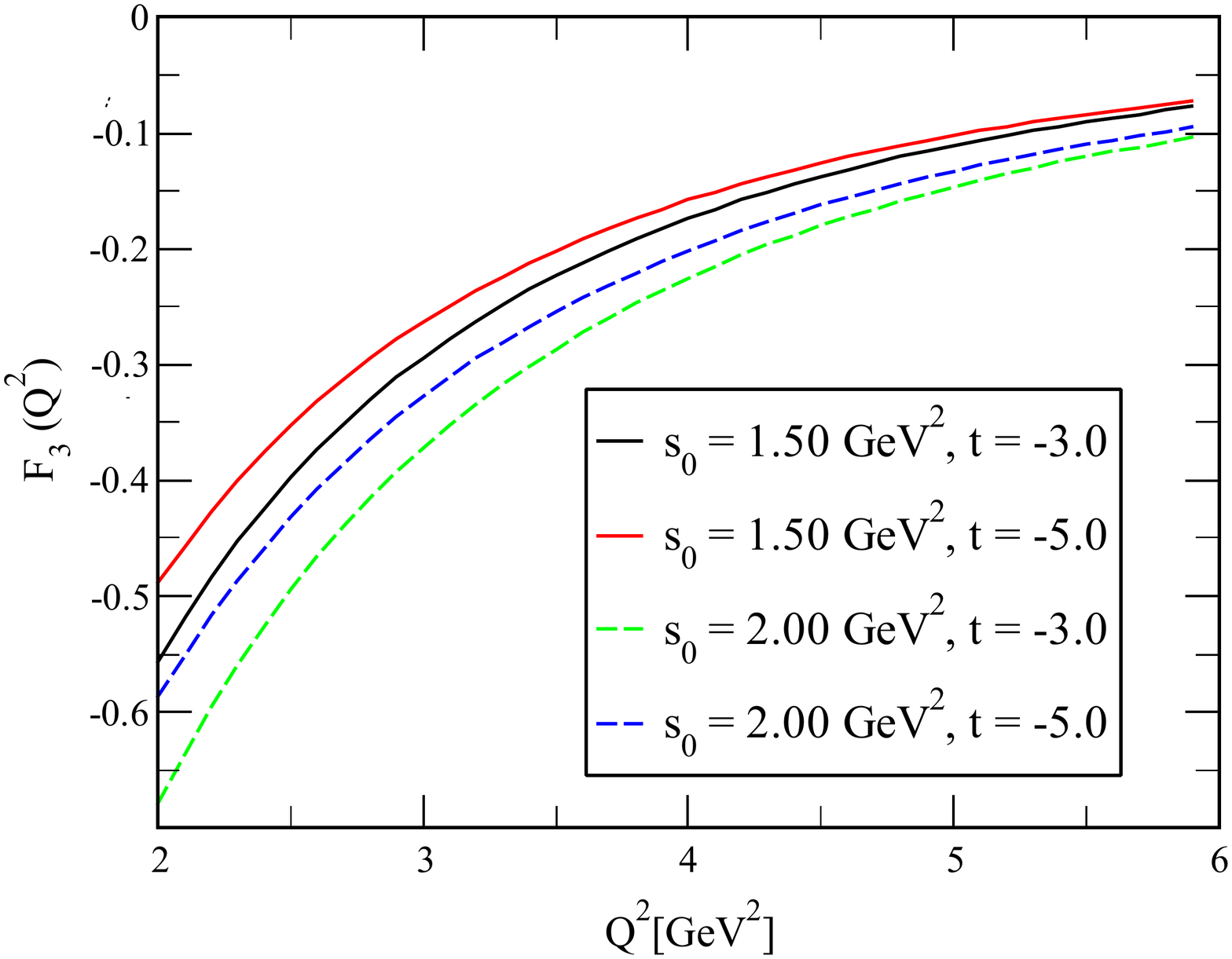}}~~~~
     \subfloat[]{\includegraphics[width=0.4\textwidth]{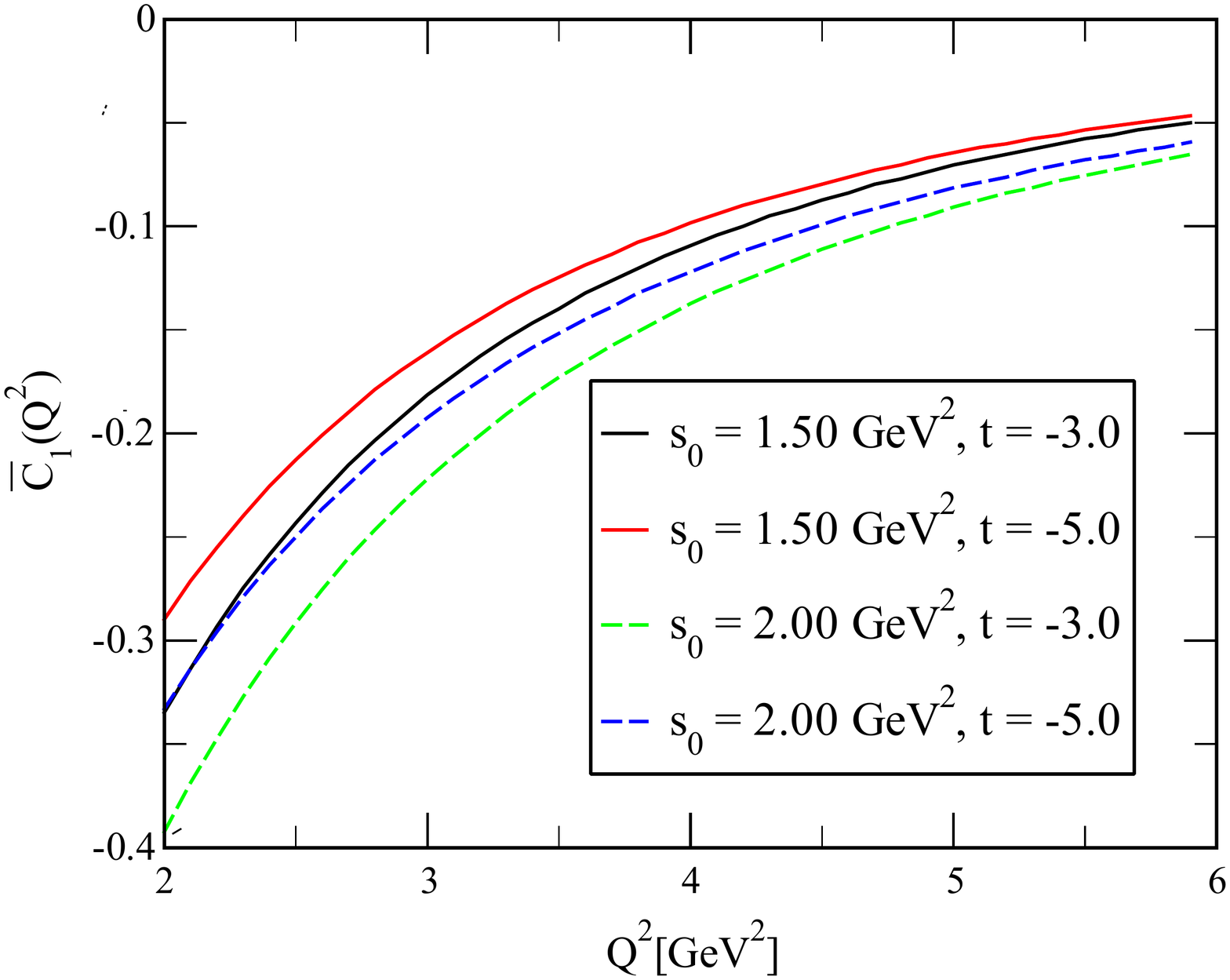}}\\
 \subfloat[]{\includegraphics[width=0.4\textwidth]{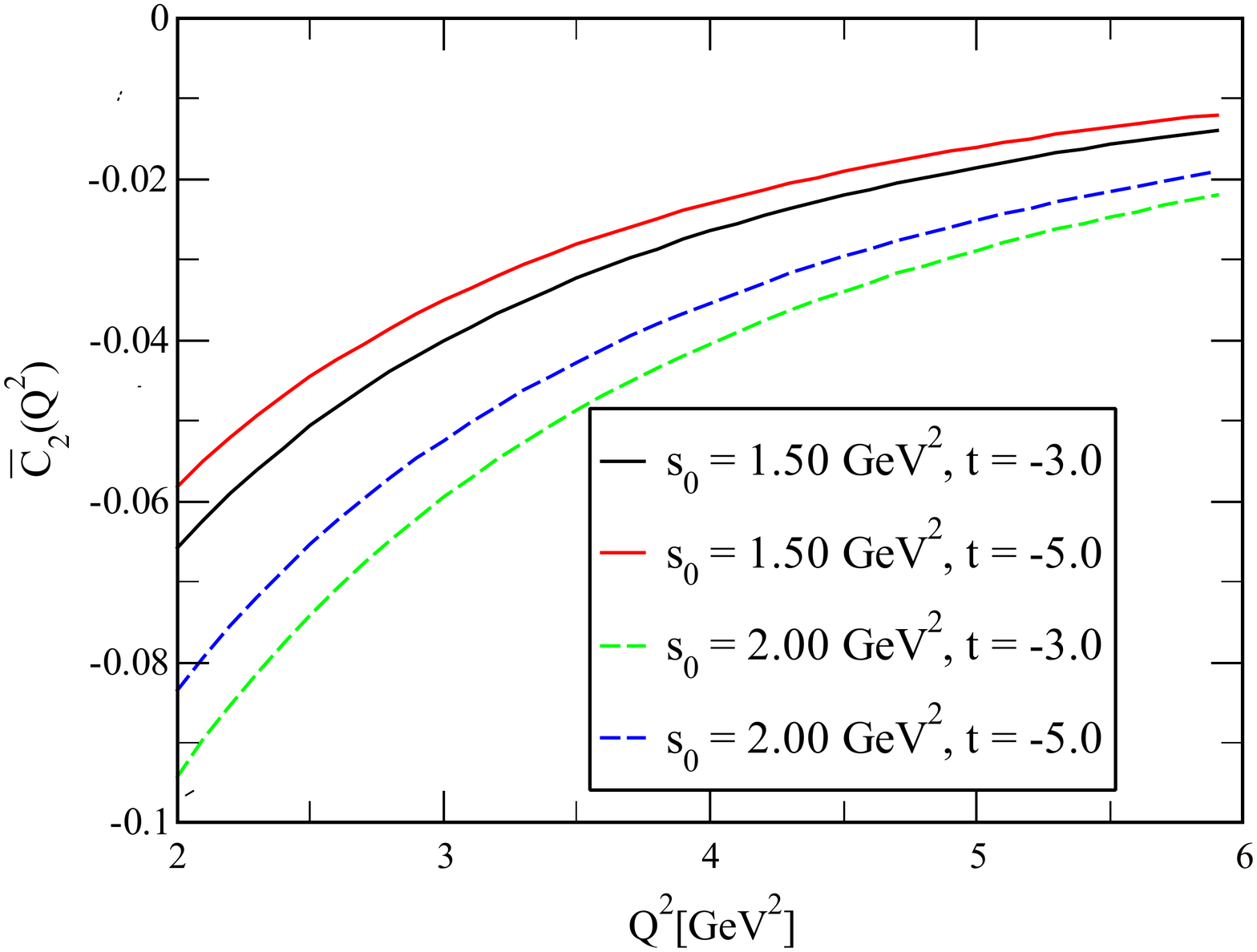}}~~~~
  \subfloat[]{\includegraphics[width=0.4\textwidth]{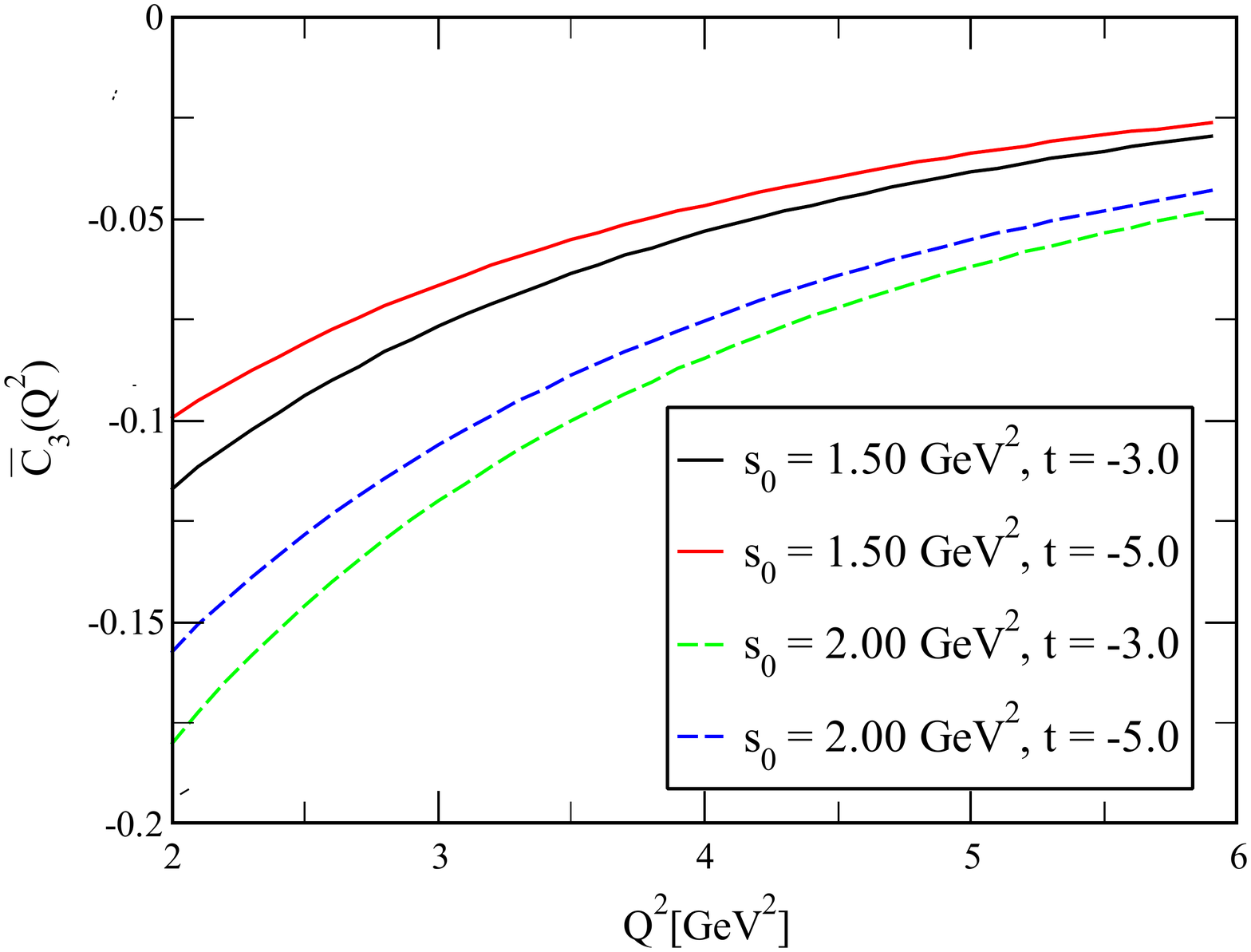}}\\
 \caption{The dependence of the $F_1 (Q^2)$, $F_2 (Q^2)$, $F_3(Q^2)$, $\bar C_1 (Q^2)$, $\bar C_2 (Q^2)$ and $\bar C_3 (Q^2)$  GFFs on $Q^2$ at fixed values of the  $s_0$,  $M^2$  and   mixing parameter $ t $.}
 \label{Qsqfigs1}
  \end{figure}

\begin{figure}[htp]
\centering
 \subfloat[]{\includegraphics[width=0.4\textwidth]{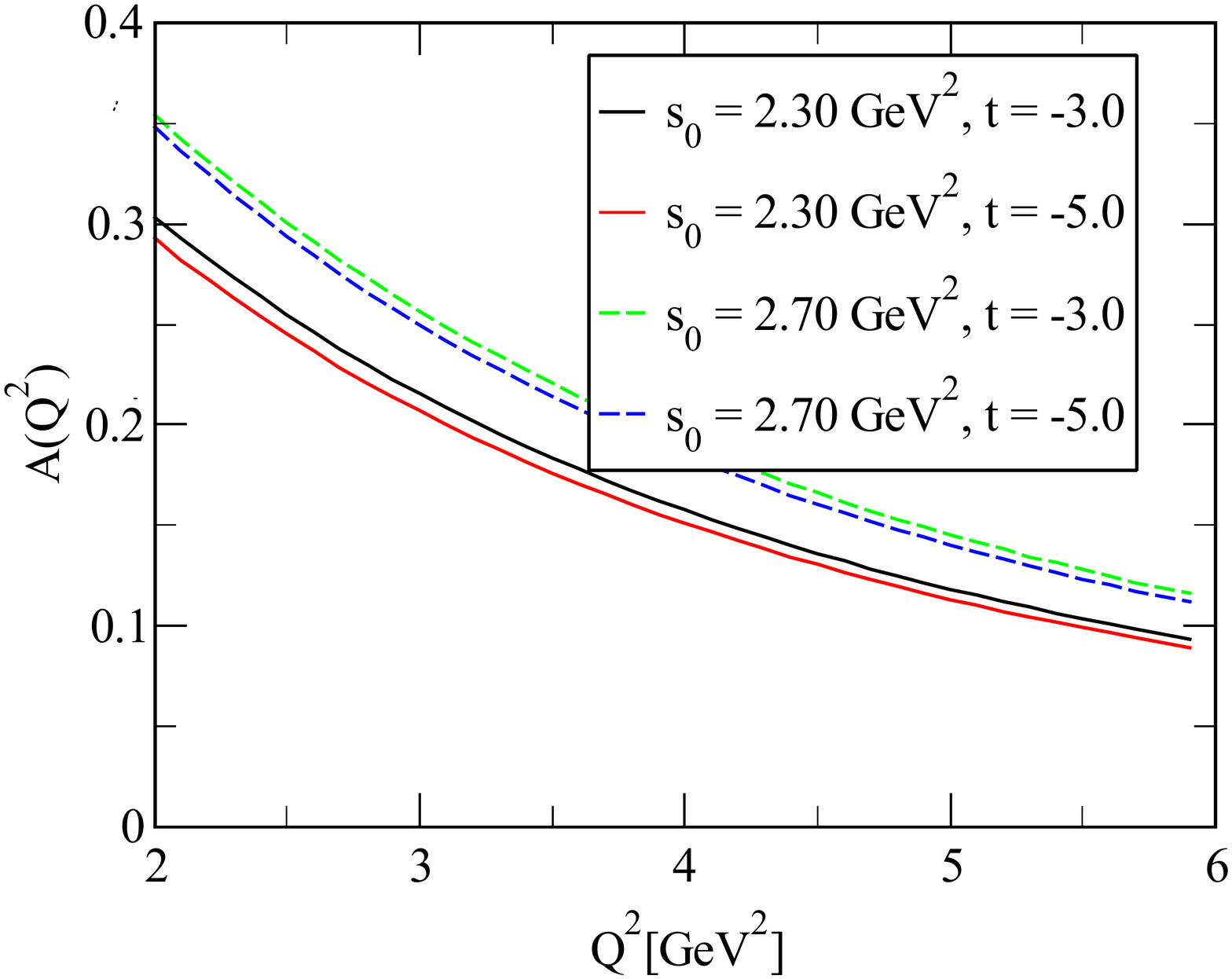}}~~~~
    \subfloat[]{\includegraphics[width=0.4\textwidth]{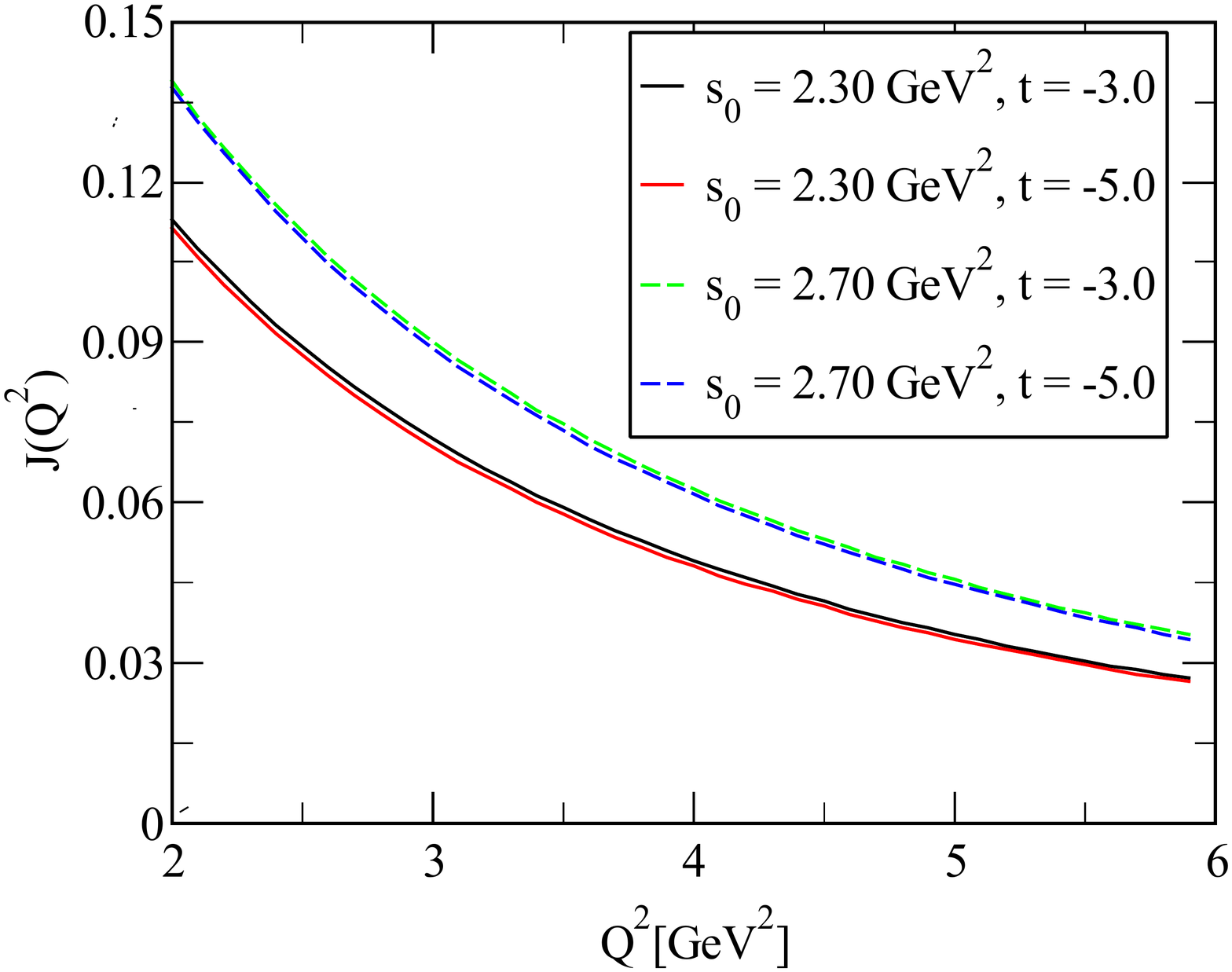}}\\
    \subfloat[]{\includegraphics[width=0.4\textwidth]{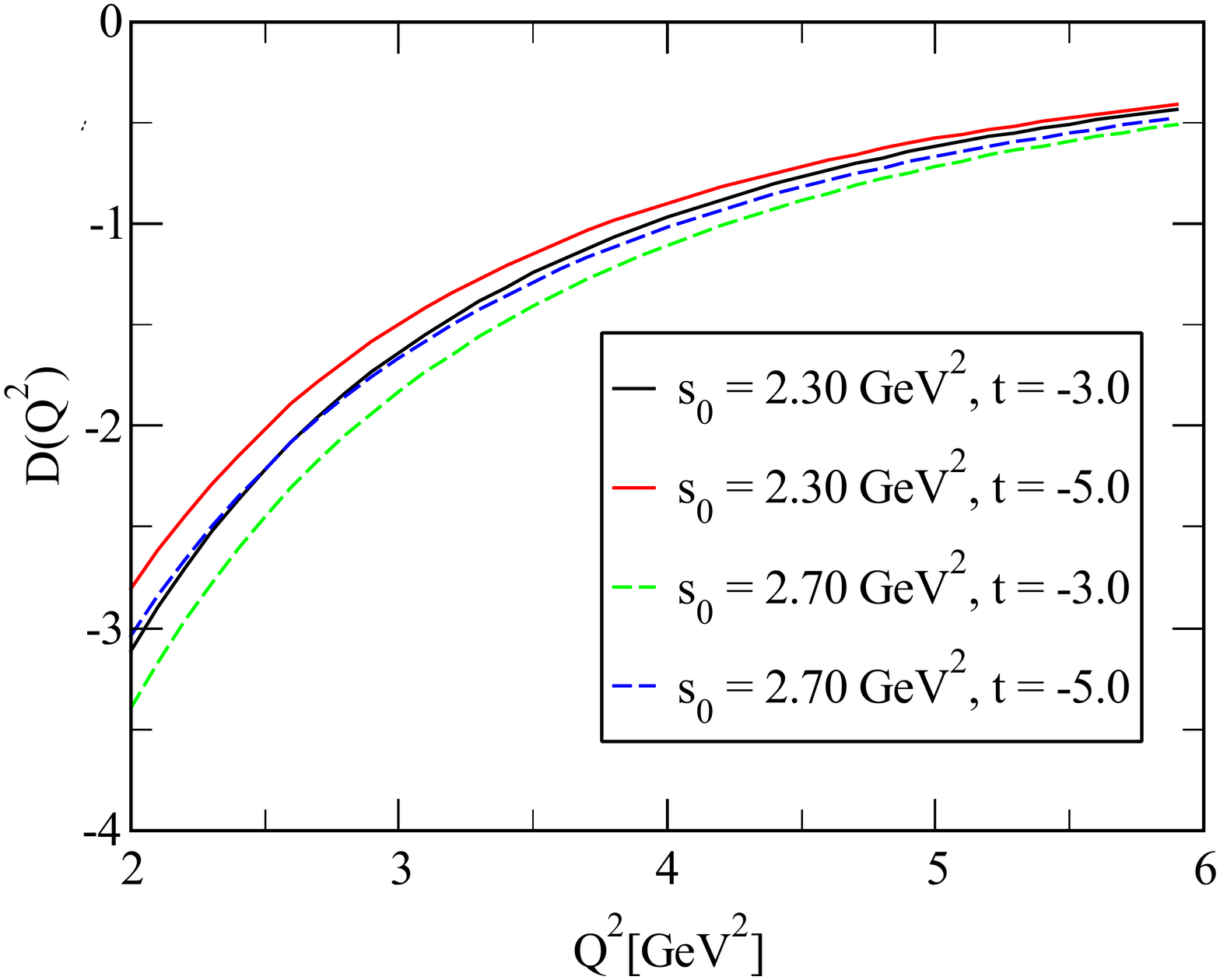}}~~~~
     \subfloat[]{\includegraphics[width=0.4\textwidth]{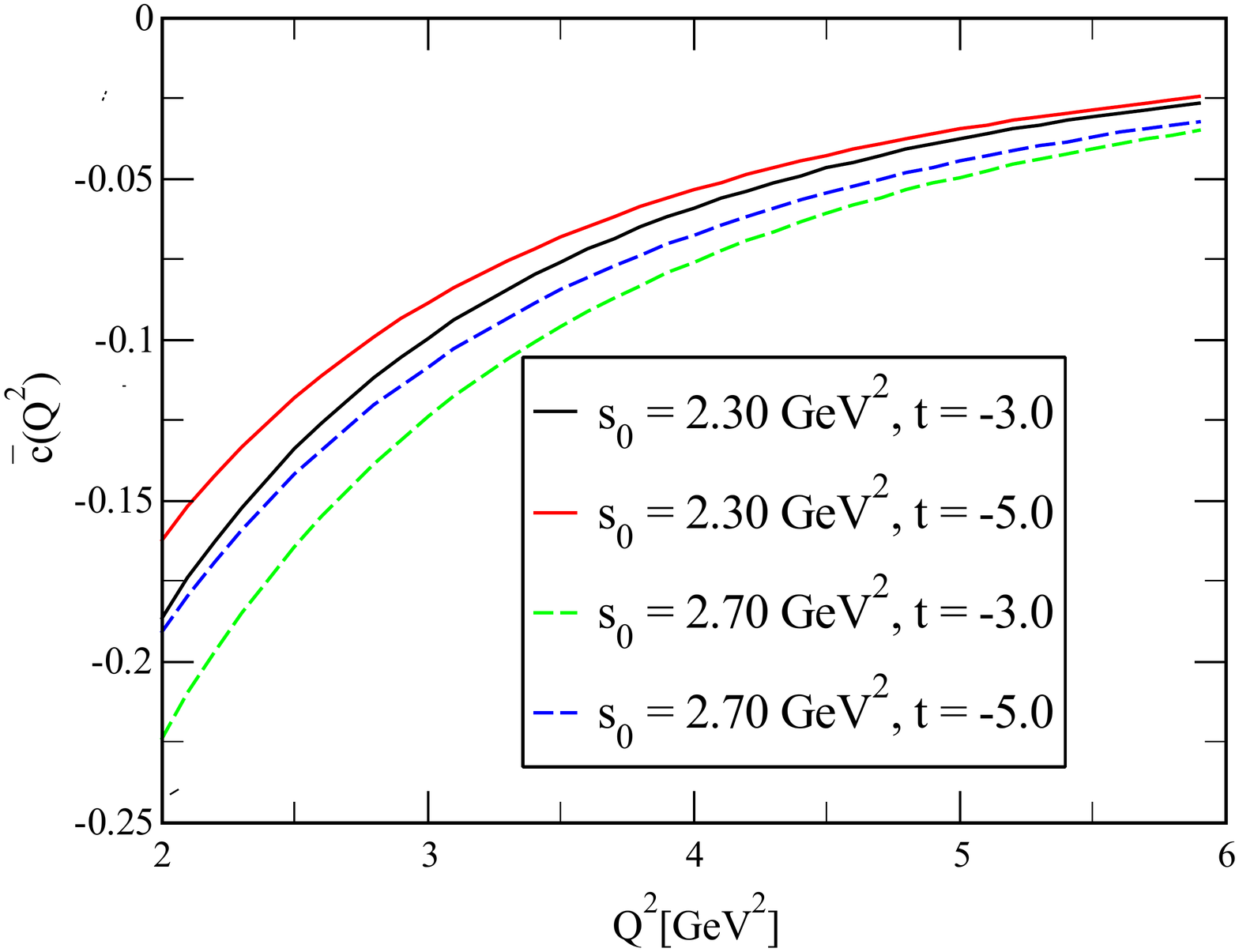}}\\
 \caption{The dependence of the $A (Q^2)$, $J (Q^2)$, $D(Q^2)$ and $\bar c (Q^2)$  GFFs on $Q^2$ at fixed values of the  $s_0$,  $M^2$  and   mixing parameter $ t $.}
 \label{Qsqfigs2}
  \end{figure}

\section{Summary and Concluding Remarks}\label{secIV}

 The FFs of hadrons due to different types of interactions are main objects, determinations of which help us get useful knowledge on various observables related to the corresponding interactions. The GFFs, which appear as a result of  interactions of hadrons with energy momentum tensor current, are of great  importance as they provide useful information on the internal structure, geometric shape, distributions of the energy and pressure as well as distribution of the strong force inside the hadrons.  In the present study, we  calculated the gravitational form factors of the excited  $N^*$ state with the quantum numbers $I(J^P)=\frac{1}{2}(\frac{1}{2}^-)$ via LCSR approach.  We  considered the quark part of the  EMT current   and used the general form of the nucleon's interpolating current  together with the DAs of $N^*$.  As both the nucleon and $N^*$ couple to the same current, the $N^* \rightarrow N$ gravitational transition form factors are entered to the calculations as the main input parameters. Hence, first we revisited  the  $N^* \rightarrow N$  transition GFFs considering the non-conservation of the quark part of the EMT current and including into analyses the six related form factors. Using the obtained results, we calculated the GFFs of  the $N^*$ excited state. We saw that, behavior of GFFs of $N^*$ in terms of $Q^2$ are well described by the  multipole fit function. Our result of $D$-term for $N^*$ is well consistent with the only prediction made by the bag model \cite{Neubelt:2019sou}.

 As a byproduct, we also calculated the  pressure and  energy density at the center of $N(1535)$ and estimated its mechanical radius using the $Q^2$ -dependent GFFs of $N^*$.  We found that the  mean mechanical radius squared of  $N^*$ state is about  $20\%$ larger than that of the nucleon. Our results may be checked via different non-perturbative methods. By the recent progresses in the experimental side, we hope that  these form factors will be measured  by near future experiments.

\bibliography{refs}

\end{document}